\documentstyle[12pt]{article}

\newcommand{\avg}[1]{\langle #1 \rangle}

\begin{document}

\title{\large{\bf NEW NEUTRINO PHYSICS WITHOUT FINE-TUNING }}
\author{C.P.~BURGESS\thanks{Talk given at Beyond the Standard Model III,
Carleton University, June 1992.}, JAMES M. CLINE\\
{\em Physics Department, McGill University, 3600 University Street,}\\
{\em Montr\'eal, Qu\'ebec, Canada, H3A 2T8.}\\[0.3cm]
and\\[0.3cm]
MARKUS LUTY\\
{\em Physics Division, Lawrence Berkeley Laboratory, 1 Cyclotron Road,} \\
{\em Berkeley, California, 94707, USA.}}
\date{{\sl August 1992 \\ McGill-92/36}}
\maketitle
\setlength{\baselineskip}{2.6ex}

\begin{center}
\parbox{13.0cm}
{\begin{center} ABSTRACT \end{center}
{\small \hspace*{0.3cm}
We show how a 17 keV neutrino, the solar neutrino problem, and the
atmospheric muon-neutrino deficit could all be the low-energy residues of the
same pattern of lepton-number breaking at and above the weak scale, with
no requirement for fine-tuning a symmetry-breaking scale at lower energies.
The required pattern of small neutrino masses turns out to be naturally
understood in this framework in terms of powers of the ratio of two high energy
scales. All cosmological and astrophysical constraints are satisfied.}}
 \end{center}

\section{Introduction}
There are several kinds of reported experimental anomalies which suggest that
new physics might be lurking in the neutrino sector. The list of titillating
phenomena includes (1) the solar-neutrino problem,\cite{snp} (2) the 17 keV
neutrino,\cite{simpson} (3) the atmospheric muon-neutrino
deficit,\cite{atmospheric} and (4) the excess high-energy electrons in
double-beta decay spectra.\cite{moe} Many of these effects remain
controversial,
with more experiments underway to determine which might be real. Until the
experimental dust settles, theorists can help by exploring the implications of
these anomalies with the goal of discovering which are consistent with one
another, potentially being signatures of the same type of underlying new
physics.

We argue here that the first three items on this list all point toward a
specific form for the neutrino mass matrix which can arise naturally from new
physics at high scales.\cite{us} The scenario to which we are led
does not include the remaining two items, although these may be separately
treated in a similar way.\cite{majorons}

A key feature of our approach is to demand that our models be ``natural'' ---
{\it i.e.} that they should involve no new symmetry breaking scales below
the weak scale. We do so because if such a large heirarchy of scales
is fine-tuned in by hand it is unstable against renormalization. Of all the
many well-understood heirarchies of scale known in nature, none have ever
been unstable in this way. But all of the ways we know for stabilizing such a
heirarchy (supersymmetry or some form of compositeness for example)
necessarily involve new particles and interactions not far above the
lower of the two scales. It is therefore hard to imagine explaining such a
heirarchy  below the weak scale without introducing new light particles which
should already have been observed.

 \section{General Constraints}
There are a number of constraints that must be satisfied by any
candidate theory of the solar neutrino problem and the 17 keV
neutrino,\cite{peccei} and these lead to several general properties that
are shared by virtually all
models. Firstly, the failure to observe neutrinoless double beta
decay implies the cancellation in this decay of the contribution of any
17 keV neutrino which enjoys a 10\% mixing with $\nu_e$. This cancellation
arises most naturally if it is enforced by a conservation law, such as the
(approximate) conservation of lepton number. In this case the 17 keV state is
(pseudo--) Dirac; {\it i.e.} it consists of a (nearly) degenerate pair of
states
which have opposite $CP$ parities. Supernova and laboratory
constraints\cite{supernova} then further suggest that these two states should
be
dominantly $\nu_\mu$ and $\nu_\tau$. The combination of lepton numbers whose
conservation can forbid double beta decay in this way is $L_e - L_\mu +
L_\tau$.

Next, a neutrino solution to the solar neutrino problem requires a fourth
neutrino state, $\nu_s$, that is approximately degenerate with $\nu_e$. The
failure to observed this new neutrino in the $Z^0$ width at LEP implies that
it must be sterile --- {\it i.e.} it cannot carry $SU(2)_W \times U(1)_Y$
quantum numbers. If this sterile neutrino were massless its renormalizable
couplings would enjoy a global symmetry, $U(1)_s$, under which $\nu_s$
transforms but all other neutrinos are neutral. This survives as
an approximate symmetry in the full theory since this mass must be very small
if it is to participate in a solution to the solar neutrino problem.
The total approximate symmetry group then becomes $G_\nu \equiv
U(1)_{e-\mu +\tau} \times U(1)_s$.

Finally, standard cosmology indicates that a 17 keV neutrino
cannot be too long-lived. Consistency with the age of the universe requires
such a neutrino to decay into relativistic daughters with a lifetime no longer
than $\tau_{17} \sim 10^{12}$ sec. Requiring
these hot decay products to not overprolong the
radiation-dominated era and so delay the onset of galaxy formation gives a
shorter bound\cite{structure} of $\tau_{17} \sim 10^6$ sec, although $\tau_{17}
\sim 10^7$ sec might actually be a good thing.\cite{goodthing} In the models to
be considered nucleosynthesis constraints are easily satisfied since all excess
light states decouple early enough to be diluted by the QCD phase transition.

Another lifetime constraint applies should the 17 keV neutrino decay into
$\overline{\nu}_e$'s. Consistency of the length of the neutrino pulse observed
from SN1987A with supernova models requires such a lifetime to lie outside
of the interval $3 \times 10^4 < \tau_{17} < 2 \times 10^8$ sec.

All of these lifetime constraints are accomodated in what follows by the decay
of the 17 keV state into a lighter neutrino plus a Goldstone boson (majoron)
that is associated with the spontaneous breaking of the approximately-conserved
lepton numbers, $G_\nu$.

\section{Models}
In order to ensure the naturalness of our models we construct them by writing
down all possible interactions that involve the required low-energy
particles and which respect the gauge and $G_\nu$ symmetries. Since all of the
neutrino effects being described arise in low-energy experiments we start
by writing down the effective lagrangian as seen at the weak scale. Any
physics of still higher scales can only affect these experiments through the
nonrenormalizable interactions they generate in this effective lagrangian.
Renormalizable models for this underlying physics are easily constructed once
the required form for the weak-scale lagrangian is known.\cite{us}

The particle content of the weak-scale theory consists of the usual
standard-model content supplemented by the sterile neutrino, $\nu_s$,
and two electroweak-singlet complex scalars, $\phi_i$. These scalars are
required in order to spontaneously break $G_\nu$, and they must be electroweak
singlets to avoid having the resulting Goldstone bosons acquire a
phenomenologically unacceptable coupling to quarks and charged leptons. The
various kinds of models we consider then differ only in their choices for the
transformation properties for $\phi_i$ under the symmetry $G_\nu$.  We present
two illustrative examples below:

\subsection{A Vacuum-Oscillation Model}
Consider first the case for which the scalar fields transform under
$G_\nu$ as $\phi_1 \sim (1/2,-1/2)$ and $\phi_2 \sim (-1/2,
-1/2)$.  Then the renormalizable lagrangian consists of the Standard Model,
a kinetic term for $\nu_s$, and kinetic and potential terms for $\phi_i$.
The lowest-dimension gauge- and $G_\nu$-invariant operators in the effective
lagrangian that can contribute to the neutrino mass matrix are
\begin{eqnarray}
\hbox{dimension 5:} \quad && \frac{g_e}{M} \, (L_e H)(L_\mu H),
\quad \frac{g_\tau}{M} \, (L_\mu H)(L_\tau H); \nonumber\\
\hbox{dimension 6:} \quad && \frac{a_j}{M^2} \, (L_j H) \, \nu_s \phi_2^2,
\quad \frac{b}{M^2} \, (L_\mu H) \, \nu_s \phi_1^2;\nonumber\\
\hbox{dimension 7:} \quad && \frac{c}{M^3} \, \nu_s\nu_s \phi_1^2 \phi_2^2;
\nonumber\\ \hbox{dimension 9:} \quad && \frac{d_{\mu\mu}}{M^5} \,
(L_\mu H)(L_\mu H) (\phi_1 \phi_2^*)^2,
\quad \frac{d_{jk}}{M^5} \, (L_j H)(L_k H) (\phi_1^* \phi_2)^2.
\end{eqnarray}
$H$ is the usual electroweak Higgs doublet and the $L$'s are the standard
left-handed lepton doublets, for which subscripts represent generation labels.
The labels $j$ and $k$ are restricted to take only the two values $e$ and
$\tau$. Appropriate factors of the heavy mass scale $M$ are included to ensure
that the remaining coupling constants are dimensionless.

Replacing the scalars with their vacuum expectation values: $\avg{H} = v = 174$
GeV, $\avg{\phi_1} = w_1$, $\avg{\phi_2} = w_2$, and defining
$g=\sqrt{g_e^2+g_\tau^2}$, gives a mass matrix of the form
 \begin{equation}
{\bf m}  = m_{17} \pmatrix{\gamma & \alpha_1 & \beta & \alpha_2 \cr
\alpha_1 & \epsilon_1 & s & \epsilon_2 \cr
\beta & s & \eta & c \cr
\alpha_2 & \epsilon_2 & c & \epsilon_3 \cr},
 \end{equation}
with $m_{17} = gv^2/M$, $\tan\theta_{17} = s/c = g_e/g_\tau$, $\alpha_j  =
(a_j w_2^2)/(g Mv)$, $\beta = (b w_1^2)/(g Mv)$, $\gamma = (c w_1^2 w_2^2)
/(g M^2 v^2)$, and $\epsilon_j, \eta = (d w_1^2 w_2^2)/(g M^4)$.

Requiring $m_{17} = 17$ keV puts the new-physics scale at $M = 1 \times 10^7
g v \approx 2 \times 10^9 g$ GeV. Since the expectation values, $v, w_1$
and $w_2$ are of order the weak scale they are much smaller than $M$. This
implies the hierarchy $\epsilon,\eta \ll \gamma \ll \alpha, \beta$.
The spectrum of neutrino masses predicted becomes: (1) a pseudo-Dirac pair of
17 keV states split by $\Delta m_h^2 = 4m_{17}^2 \beta \alpha_2'$, and
(2) a pseudo-Dirac pair of light neutrino states with $m_\ell = m_{17}
\alpha_1'$, and $\Delta m_\ell^2 = m_{17}^2 \gamma \alpha_1'$. In these
expressions the coefficients $\alpha_i'$ are defined by: $\alpha_1' = c
\alpha_1
- s \alpha_2$ and  $\alpha_2' = c \alpha_2 + s \alpha_1$.

There are two majorons in this model, $\chi_1$ and $\chi_2$, which can be
thought of as the phases of the fields $\phi_1$ and $\phi_2$, respectively.
The lifetime for the decay $\nu_h \to \nu_\ell + \chi_i$ in this model is
\begin{equation}
\tau_{17} = \frac{16\pi}{m_h^3} \left( \frac{\alpha_2'^2}{w_2^2} +
\frac{\beta^2}{w_1^2} \right)^{-1}
\end{equation}

A typical choice for the couplings is given by $w_1, w_2 \sim 3v$, $a_j = b
= c = g_\tau = 1$ and $g_e = 0.1$.  In this case $\alpha', \beta \sim 10^{-6}$,
 $\gamma \sim 10^{-12}$, and the neutrino masses become $m_\ell \sim 0.01$ eV,
$\Delta m_\ell^2 \sim 10^{-10}$ eV${}^2$, and $\Delta m_h^2 \sim 10^{-3}$
eV${}^2$. There is maximal (45${}^o$) mixing within each of these pseudo-Dirac
pairs. These numbers also imply a lifetime of $\tau_{17} \sim 10^9$ sec.

It is striking that the hierarchies $m_h / v$ and $m_\ell / m_h$, as well as
$\Delta m_h^2 / m_h^2$ and $\Delta m_\ell^2 / m_\ell^2$ are all explained here
by the largeness of $M$ relative to $v$, $w_1$ and $w_2$. $\Delta m_h^2$ is
near
the experimental upper limit and in the range required to account for the
atmospheric neutrino anomaly, and $\Delta m^2_\ell$ falls naturally into the
correct range for ``just-so'' vacuum oscillations.

These numbers easily satisfy the phenomenological constraints
with two provisos. Although they can account for the atmospheric neutrino
deficit, they do so with maximal $\nu_\mu - \nu_\tau$ oscillations. They
therefore require that either the atmospheric neutrino deficit itself, or the
most recent IMB constraints coming from the flux of higher-energy upward-coming
muons, must disappear. The 17 keV lifetime is also long compared to the
structure-formation bound although, as may be seen in the following section, it
can be shortened with alternative choices for the dimensionless couplings.

\subsection{An MSW Model}
A slightly different choice for the quantum numbers for $\phi_i$ produces
a model with resonant MSW oscillations. In this case the singlet scalar fields
transform under $G_\nu$ as $\phi_1 \sim ( -1/2, -1/2)$ and $\phi_2 \sim
(0,-2/3)$. The lowest-dimension gauge- and $G_\nu$-invariant operators
that contribute to the neutrino mass matrix are in this case
\begin{eqnarray}
\hbox{dimension 5:} \quad && \frac {g_e}{M} \, (L_e H)(L_\mu H),
\quad \frac {g_\tau}{M} \, (L_\mu H)(L_\tau H); \nonumber\\
\hbox{dimension 6:} \quad && \frac {a_j}{M^2} \, (L_j H) \, \nu_s \phi_1^2,
\quad \frac {c}{M^3} \, \nu_s\nu_s \phi_2^3.
 \end{eqnarray}
Contributions to
the remaining terms in the neutrino mass matrix are further suppressed relative
to these by additional powers of $M^{-1}$.

In this case the mass matrix again takes the form of Eq. (2) with $\alpha_j
= (a_j w_1^2)/(g Mv)$, $\gamma = (c w_2^3)/(g M v^2)$, and $\beta, \epsilon,
\eta \ll \alpha, \gamma$. This implies the light neutrino states have masses
$m_{\ell\pm} = (m_{17}/2) \; \left[ \sqrt{\gamma^2 + 4{\alpha_1'}^2} \pm \gamma
\right]$, and $\Delta m_\ell^2 = m_{17}^2 \gamma \sqrt{\gamma^2 +
4{\alpha_1'}^2}$. The mixing angle between the two light states works out to be
$\sin^2 2\theta_\ell = 4{\alpha_1'}^2/(\gamma^2 + 4{\alpha_1'}^2)$. The 17
keV state is negligibly split in this model.
 MSW oscillations of the light states can be accomodated if we choose
$g = 1$, $a_1 = 0.2$, $a_2 = 1$, and $c = 1$, for which $\alpha'_1 = 1
\times 10^{-8}$, $\alpha'_2 = 1 \times 10^{-7}$, $\gamma = 1 \times 10^{-7}$,
$\Delta m_\ell^2 = 3 \times 10^{-6}$ eV${}^2$, and $\sin^2 2\theta_\ell = 4
\times 10^{-2}$. These values do give trouble with the structure-formation
 bound, however, since they give a lifetime for the 17 keV state of $\sim
10^{10}$ sec.

Shorter lifetimes as well as the atmospheric neutrino anomaly can be
accomodated
by adding a third electroweak singlet scalar transforming under $G_\nu$ as
$\phi_3 \sim (1/2, -1/2)$ since this allows the additional dimension 6
operator:
\begin{equation}
\frac b{M^2} \; (L_\mu H) \nu_s \phi_3^2.
\end{equation}
Resolution of the atmospheric neutrino anomaly requires $w_3$ near the weak
scale. For example, $w_1 = v / 2$, $w_2 = v$,
$w_3 = 30v$, $g = 0.1$, $a_1 = 0.1$, $a_2 = 0.01$, $b=1$, and $c = 0.01$,
gives the MSW effect, atmospheric neutrino oscillations, and a neutrino
lifetime of $\tau_{17} = (16\pi w_3^2)/(m_h^3\beta^2) \sim 2 \times 10^{3}$
sec.

\section{Experimental Implications}

There are several ways in which the class of models we have discussed might be
experimentally probed. The most obvious way is to confirm or disprove
the 17 keV neutrino and the solar neutrino problem. Solar neutrino
oscillations are also predicted to be into a sterile component --- a prediction
that is potentially detectable at SNO. If the atmospheric neutrino persists, it
must be explained in this picture using maximal $\nu_\mu - \nu_\tau$
oscillations, which would imply that the IMB results on the upcoming muon flux
cannot survive. Since the 17 keV lifetime likes to be long in this scenario, it
may have interesting applications for galaxy formation.

\def\prl#1#2#3{{\it Phys. Rev. Lett.} {\bf #1} (19#2) #3}
\def\plb#1#2#3{{\it Phys. Lett.} {\bf B#1} (19#2) #3}
\def\npb#1#2#3{{\it Nuc. Phys.} {\bf B#1} (19#2) #3}
\def\etal{{\it et.al. \/}}

\bibliographystyle{unsrt}

\end{document}